\documentstyle[12pt]{article}
\begin{document}
\title{CP violation in the annihilation $e^{+}~e^{-} \rightarrow t~\bar{t}$
       within the Model-III of the two-Higgs-doublet extension}

\vspace{-3mm}
\author{{ Chao-Hsi Chang$^{a,c}$, Han Liang$^{b}$, Jiang Yi$^{b}$,
Li Xue-Qian$^{a,c,d}$}\\
{Ma Wen-Gan$^{a,b,c}$, Zhou Hong$^{b}$ and Zhou Mian-Lai$^{b}$}\\
{\small $^{a}$CCAST (World Laboratory), P.O.Box 8730, Beijing 100080, China} \\
{\small $^{b}$Modern Physics Department, University of Science and}\\
{\small Technology of China, Anhui 230027, China}  \\
{\small $^{c}$Institute of Theoretical Physics, Academia Sinica,
        P.O.Box 2735, Beijing 100080, China.}  \\
{\small $^{d}$Department of Physics, Nankai University, Tianjing
        300071, China.}}

\date{}
\maketitle
\vspace{-8mm}

\begin{abstract}
The CP-violating effects are studied in the
electron-positron annihilation to produce $t\bar{t}$ pair
within the Model-III of the two-Higgs-doublet extensions,
which allows flavor changing neutral currents via Higgs exchanges.
Complete analytical expressions for
the form factors of the top quark electric and weak dipole moments
induced in the model are presented. Several observables sensitive to
the CP violation are calculated. The dependences of the
CP-violating effects on Higgs boson masses and the possible
phase angles are discussed.
We find that the CP-violating observables can be
of the order of $10^{-4}$ in general, but may reach to
$10^{-3}$ for the most favorable parameters.

\vspace{4mm}
\end{abstract}

\vfill \eject


\noindent{\large\bf I. Introduction}

Symmetries play important role in physics. In the real physical world,
some of the symmetries are exact and some are broken.
The symmetry CP was found to be broken in neutral kaon system
in 1964 by Christenson, Cronin, Fitch and Turlay\cite{Christ}.
Since the discovery on CP violation, more than 30 years have passed.
Various models have been proposed to explain the observed
CP violation in $K^{0}-\bar{K}^{0}$ mixing.
It is well known that the Kobayashi-Maskawa(KM) mechanism
in the Standard Model(SM) framework is consistent with all experimental data.
Whereas to explain the exist experimental data, there is still some room for
certain extensions of the SM. Especially, it seems that to solve
the so-called baryon genesis problem of the universe, the strength
of the CP violation due to the KM mechanism is not strong enough and it
indicates some new source(s) of CP violation else is required.
Therefore the problem on the source(s) of CP violation
should be considered being still open. In the literatures\cite{Lee,Wu},
it is pointed out that the extended Higgs sector in the two-Higgs-doublet
model(THDM) may open many different and interesting CP violating
sources. Indeed, if CP violation is really related to
the Higgs sector of a SM extension, such as in the THDM-III
which we are considering
in this work, the verification on the CP violation
source is crucial not only for the problem itself,
but also for understanding the electroweak symmetry
breaking. Moreover as mentioned above, it also must be interesting for
cosmology study, i.e., opening some solutions to the baryon-genesis problem
in the universe.

In the commonly discussed THDM, the natural flavor
conservation(NFC) condition is invoked to assure the
absence of flavor changing scalar interactions(FCSI)
by certain symmetry. Then the THDM with the NFC condition is
subdivided into two modes, namely Model-I and Model-II.
In Model-I, both the up and down type quarks achieve
their masses from the same Higgs doublet, and in Model-II the quarks do so
from different doublets. In fact, to compare with experimental data,
it is not so necessary to put a symmetry to dictate the NFC condition,
whereas, there is another possible mode of the THDM, which is popularly called
as Model-III now. In this model instead of placing the NFC constraints
by a discrete symmetry, the Yukawa couplings for FCSI are introduced
as usual done, i.e., in the meantime, they also generate the fermions' masses
with the nonzero vacuum expectation value of the Higgs. The consequence
of the way to introduce the Yukawa couplings for the THDM-III, is that
the coupling will be proportional to the masses of the coupled fermions.
Considering the fact that the most fundamental fermions
in the SM except top-quark have a small mass, the FCSI in the THDM-III
at the tree level would not exist substantially at low mass scales.
As we know that two years ago the CDF and D0 collaborations found the
top quark has a so large mass(i.e., now the
world average value: $m_{t} = 175.6 \pm 5.5~GeV$)\cite{CDF}.
Therefore it is believed that careful study on the processes relevant
to the top quark would be the most possible approach to new physics.
As the heavy top quark will strongly suppress the effects
of its mixing with other generations in the SM, the GIM mechanism of
unitary constraints leads to very small CP-violation effects, although
it makes the greater CP effects due to KM mechanism in B meson decays
and the B meson systems than those in K systems.
Hence any observation of the CP violation in top quark production
and decays would give interesting informations, because it may relate to
the physics beyond the SM including the THDM-III. In addition, the lifetime
of top quark is so short that there is no enough time to make hadrons
before its weak decay, the `phase information' of the produced top quark,
which contains the CP signature, will be kept in its decay products
very well without `disturbing' by hadronization. Therefore to observe
the CP violation in top quark production makes sense and becomes accessible.

Previously, motivating CP violation in the top production
has been extensively investigated\cite{Bern}\cite{Christ1}\cite{Schmidt}.
In most of these works, the effects from the electric and weak
dipoles and the CP-violating observables in the process $e^{+} e^{-}
\rightarrow t \bar{t}$ are investigated in a model-independent way.
Some works are contributed to the CP-violating effects induced by the
Higgs sector in the THDM without FCSI's and the MSSM in the process.

In this paper we concentrate on the CP violation effects in
the process of the top quark pair production at $e^{+} e^{-}$ colliders,
which are originated from the THDM-III.
We organize the paper as follows: In section 2, we will briefly
outline the THDM-III for self containment and give the relevant
vertices of Yukawa couplings involving CP-violating phases.
In section 3, the complete electric and weak dipole moment form factors
of the top quark within the THDM-III are calculated. In section 4,
some CP violating sensitive observables for the process
$e^{+} e^{-} \rightarrow t \bar{t}$ are evaluated and discussed. Finally
a short summary is given.

\vskip 5mm
\noindent{\large\bf 2. The third type of the two Higgs doublet model(THDM-III)}

Generally in the third type of the two-Higgs-doublet model, the up-type
and down-type quarks couple simultaneously to more than
one scalar doublet. We consider the THDM-III has two scalar $SU(2)_{w}$
doublets, $\phi_{1}$ and $\phi_{2}$:
$$
\phi_{1}= \left( \begin{array}{l} \phi_{1}^{+} \\
          \phi_{1}^{0} \end{array} \right) ,
~~\phi_{2}= \left( \begin{array}{l} \phi_{2}^{+} \\
          \phi_{2}^{0} \end{array} \right),
\eqno{(2.1)}
$$
with Lagrangian
$$
{\cal L}_{\phi}=D^{\mu}\phi^{\dagger}_{1}D_{\mu}\phi_{1}+
               D^{\mu}\phi^{\dagger}_{2}D_{\mu}\phi_{2}-
               V(\phi_{1},\phi_{2}),
\eqno{(2.2)}
$$
where $V(\phi_{1},\phi_{2})$ is the general potential which is consistent
with the gauge symmetries. Since there is no global symmetry that
distinguishes the two doublets in the THDM-III, without loss of generality
we assume $v_{2}=0,~ \beta=\pi$ with $\tan\beta\equiv v_2/v_1$. Then we have
$$
<\phi_{1}>= \left( \begin{array}{l} 0 \\
            v/\sqrt{2} \end{array} \right) , ~~<\phi_{2}>=0
\eqno{(2.3)}
$$
where $v \simeq 246~GeV$. Three of the components of $\phi_{1}$
become the longitudinal components of the $W^{\pm}$ and $Z^{0}$ bosons,
and the physical spectrum contains also the charged scalar bosons $H^{\pm}$,
the neutral scalars $h^{0}$ and $H^{0}$ and the pseudoscalar $A^{0}$.

$$
\begin{array}{l}
H^{0}=\sqrt{2}[(Re\phi_{1}^{0}-v)\cos{\alpha}+Re\phi_{2}^{0}\sin\alpha],\\
h^{0}=\sqrt{2}[-(Re\phi_{1}^{0}-v)\sin{\alpha}+Re\phi_{2}^{0}\cos\alpha],\\
A^{0}=\sqrt{2}(-Im\phi_{2}^{0}), \\
H^{\pm}=-\phi_{2}^{\pm}.
\end{array}
\eqno{(2.4)}
$$
where $\alpha$ is the mixing angle of the CP-even neutral Higgs bosons.
The physical charged Higgs state is orthogonal to $G^{\pm}$. Then
the charged Goldstone bosons $G^{\pm}$ are
$$
\begin{array}{l}
G^{\pm}=-\phi_{1}^{\pm}.
\end{array}
\eqno{(2.5)}
$$
In this model there are still six free parameters which include the masses of
the five neutral and charged Higgs bosons($m_{h^{0}},~m_{H^{0}},~m_{A^{0}},
~m_{H^{\pm}}$) and the mixing angle $\alpha$. The Yukawa couplings to quarks
are\cite{Luke},
$$
{\cal L}^{Q}_{Y}=\lambda^{U}_{ij}\bar{Q_{i}}\tilde{\phi_{1}}U_{j}+
                \lambda^{D}_{ij}\bar{Q_{i}}\phi_{1}D_{j}+
                \xi^{U}_{ij}\bar{Q_{i}}\tilde{\phi_{2}}U_{j}+
                \xi^{D}_{ij}\bar{Q_{i}}\phi_{2}D_{j}+h.c..
\eqno{(2.6)}
$$
In above equation we define $\tilde{\phi}_{1,2}=i\tau_{2}\phi^{\ast}_{1,2}$,
$U_{i}$ and $D_{i}$ are the SU(2) doublets of left-handed quark states and
$Q_{i}~(i=1,2,3)$ are the SU(2) singlets of right-handed quark states.
In the first two terms in Eq.(2.6) involving the $\phi_{1}$ doublet
generate the quark masses, while $\phi_{2}$ appearing in the following
two terms is responsible for the new interactions. With the usual
manipulations $\lambda^{U}$ and $\lambda^{D}$ can be expressed in terms
of the mass matrices $\frac{\sqrt{2}M^{U}}{v}$ and
$\frac{\sqrt{2}M^{D}V_{KM}^{\dag}}{v}$, where $V_{KM}$ is the
Kobayashi-Maskawa(KM) matrix. $\xi^{U}_{ij}$ and
$\xi^{D}_{ij}$ are the $3 \times 3$ matrices which determine the
strengths of the flavor changing neutral scalar vertices. The $\xi$s are
all free parameters and can only be constrained by the
observation of experiments relating to FCSI.
The CP violation can be induced in this model by having complex $\xi$s
and the phase angle in KM matrix. In order to apply the THDM-III to our
specific process, we have to employ some definite ansatz on the $\xi^{U}_{ij}$
and $\xi^{D}_{ij}$. Here we use the Cheng-Sher Ansatz (CSA)\cite{chengs}
and the Yukawa couplings can be parametrized as
$$
|\xi^{U,D}_{ij}|=g \frac{\sqrt{m_{i}m_{j}}}{m_{W}} |\lambda^{U,D}_{ij}|.
\eqno{(2.7)}
$$
In principle the parameters $\lambda^{U,D}_{ij}$ need to be constrained by
available phenomenology. Due to the fact that the factor
$\frac {\sqrt {m_im_j}}{m_W}$ for light quarks is quite small,
the observations of the FCSI of the THDM-III
for light quarks are quite difficult,
so far the substantial constraints are lack i.e.,
the $\lambda^{U,D}_{ij}$ mixing parameters are not well
constrained. In order to simplified our analysis and
have a possible and definite CP violation effect,
we will take $|\lambda^{U,D}_{ij}|=\frac{1}{\sqrt{2}}$,
which may make a comparison of $\lambda^{U,D}_{ij}$ to
the usual gauge couplings of $SU(2) \times U(1)$ in the SM. Whereas
in general $\lambda^{U,D}_{ij}$ could be complex so as to
induce an additional CP violation source. Thus we write
$\lambda^{U,D}_{ij}$ as
$$
\lambda^{U,D}_{ij}=\frac{1}{\sqrt{2}} e^{\theta^{U,D}_{ij}}.
\eqno{(2.8)}
$$
where $\theta^{U,D}_{ij}~(i,j=1,2,3)$ are the phase angles of the complex
parameters $\lambda^{U,D}_{ij}$. We denote $\theta^{U}_{33}=\theta_{1}$
and $\theta^{D}_{33}=\theta_{2}$, and try to concern the CP violation
caused by them in our calculation.

\vskip 5mm
\noindent{\large\bf 3. CP-violating dipole couplings of top quark}

In the process $e^{+} e^{-} \rightarrow t \bar{t}$ the CP-violating
effects come from the electric and weak dipole moments of top quark. In this
section we analyze the THDM-III contributions to electric and weak dipoles
of the top quark. In the THDM-III, additional complex couplings can
be introduced and lead to CP violation in top quark physics.
Generally it is assumed that the couplings $-ie\Gamma^{\gamma,Z}_{\mu}$,
which are the interactions between the top quark and $\gamma$, $Z^0$.
The couplings include CP violation phases and have the form factors as:
$$
\Gamma^{j}_{\mu}=V^{j}\gamma_{\mu}+A^{j}\gamma_{\mu}\gamma_{5}+
    \frac{d_{j}}{2 m_{t}} i\gamma_{5}(p_{t}-p_{\bar{t}})_{\mu},~~~~
    (j=\gamma, Z),
\eqno{(3.1)}
$$
with
$$
V^{\gamma}=\frac{2}{3},~~~A^{\gamma}=0,~~~
V^{Z}=\frac{1}{4\sin{\theta_W}\cos{\theta_W}} (1-\frac{8}{3}
      sin^{2}\theta_W),~~~A^{Z}=-\frac{1}{4\sin{\theta_W}\cos{\theta_W}}.
\eqno{(3.2)}
$$
In Eq.(3.1), the $d_{\gamma}$ and $d_{Z}$ are called the electric
and weak dipole moment form factors, respectively.
In figure (1) we represent the one-loop Feynman diagrams of the
vertices which contribute to
the electric and weak dipole form factors of the top quark.
In our consideration we ignore the couplings of the quarks in the first
and second generations to Higgs and Goldstone bosons, because of their
very light masses. One can expect that if the Higgs boson
masses are not much heavier than the top quark, there would be a large
values for electric and weak dipoles induced by the THDM-III one-loop
corrections. The relevant Feynman rules of the interactions with
complex couplings are given in figure (2). The explicit expressions
of the electric and weak dipole form factors of top quark in the framework
of the THDM-III are listed in Appendix.

The size of the dipole moment form factors $d_{\gamma}$ and $d_{Z}$
depends strongly on the phases of the parameters $\xi$s in the
THDM-III. The electric dipole observations on neutron and other
experiments should put constraints on the phases of $\xi$s. In the
calculation on the neutron dipole it is known that
there are some theoretical uncertainties for the neutron struction of
quarks, thus at present calculation we put the experimental constraint
from neutron dipole aside and make more general analysis for the
CP-violation effects in top quark production.
In the numerical evolution we take the input parameters
as $m_b=4.5$ GeV, $m_Z=91.187$ GeV, $m_W=80.33$ GeV, $m_t=175$ GeV,
$G_F=1.166392\times 10^{-5} (GeV)^{-2}$ and $\alpha(0)=\frac{1}{137.036}$
\cite{Barn}, here we ignore the running of $\alpha$ which can only
give rise to minor changes to our results.
The real and imaginary parts of the $d_{\gamma}$ and $d_{Z}$
as the functions of the energy of the center mass system (CMS) of the incoming
electron and positron, are plotted in Fig.3(1). There we take the
mass values of Higgs bosons as $m_{h^{0}}=100~GeV$, $m_{H^{0}}=200~GeV$,
$m_{H^{\pm}}=150~GeV$ and $m_A=250~GeV$, the mixing angle of the neutral
Higgs bosons as $\alpha=\frac{\pi}{6}$, the CP-violating phase
angles as $\theta_1=\frac{\pi}{4}$ and $\theta_2=\frac{\pi}{2}$.
The absolute values of both the real and imaginary parts of the dipoles
are enhanced at vicinity of the energy region of
the top pair production threshold and
tend to vanish when the center-of-mass energy $\sqrt{s}$ is getting larger.
In Fig.3(2), Fig.3(3) and Fig.3(4), the values of electric and weak dipoles
as the functions of the masses of the Higgs bosons $m_{h^{0}}$, $m_{H^{\pm}}$
and $m_{A}$ are depicted, respectively, with $\sqrt{s}=500~GeV$ and the
other parameters being the same as those in Fig.3(1). The dependences of
the quantities of the dipoles on the masses of $h^0$ and $A^0$ Higgs bosons
are clear, but the dipoles do not depend on the mass of $H^{\pm}$ very much.
In Fig.3(2) the curves for the real and imaginary parts of weak dipole
are raised quantitatively at the position of $m_{h} \sim 400~GeV$. That is
because of the resonance effects contributed by the diagrams Fig.1(6,7,8,9)
with the condition $\sqrt{s} = m_{h}+m_{Z}$ being satisfied.
There is a small spike on each of the four curves in Fig.3(3) at the position
of $m_{H^{\pm}}=m_t+m_b \sim 180~GeV$ due to the resonance effect coming from
the one-loop diagrams in Fig.1(5). The dependences of
the dipoles on the mass of $H^0$ are not shown in figure, since the features
of these dependences are similar with that in Fig.3(2) due to the similar
Yukawa couplings with quarks. The electric and weak dipoles as the functions
of the CP-violating phase angles $\theta_{1}(\theta^{U}_{33})$ and
$\theta_{2}(\theta^{D}_{33})$
are shown in Fig.3(5) and Fig.3(6), respectively. In each of these two
figures we take the input parameter values as before and assume the phase
angles $\theta_{2}$ and $\theta_{1}$ to be zero in Fig.3(5) and Fig.3(6),
respectively. We can see that the
dipoles mainly come from the phase angle $\theta_1$ and take maximal absolute
values when $\theta_{1}$ has the values around $125^{\circ}$ and $235^{\circ}$.
The contributions to dipoles from the phase angle $\theta_2$ are two
order smaller than from the phase angle $\theta_1$. That can be
understood because the CP effect from the complex phase angle $\theta_{1}$
is in proportion to top quark mass, whereas the effect from the phase angle
$\theta_{2}$ is in proportion to bottom quark mass, which can be read from
the coupling Feynman rules shown in Fig.2.

\vskip 5mm
\noindent{\large\bf 4. CP violating observables in the process
               $e^{+} e^{-} \rightarrow t \bar{t}$}

At the future electron-positron collider CP violation in the $t \bar{t}$
pair production can be searched by various means\cite{Bern}\cite{Schmidt}.
In this section, we present some CP violation observables precisely and
then discuss the behaviors of CP violation effects due to the electric and
weak dipoles. The presence of electroweak dipole of the top quark will
induce the CP violation in the process $e^{+} e^{-} \rightarrow t \bar{t}$.
It leads asymmetry in the polarizations of the top quark and top anti-quark.
These polarizations can be determined by analyzing the energy and angular
distributions of the charged lepton(antilepton) particles produced by the
top quark(top anti-quark) sequential leptonic decay. The analytic expressions
for these distributions and the CP-violating angular asymmetries ($A_{ch}
(\theta_0)$, $A_{FB}(\theta_0)$) have been obtained
by Poulose and Rindani\cite{Poul}. In those expressions the CP-violating
effects in the consequent decays of top quark and top antiquark are ignored.
The total lepton charge asymmetry with a angle cutoff $\theta_0$ of the
charged lepton is defined as:
$$
{\cal A}_{ch}(\theta_0)=\frac{\int_{\theta_0}^{\pi-\theta_0} d\theta_{l}
  \left( \frac{d\sigma^{+}}{d\theta_l}-\frac{d\sigma^{-}}{d\theta_l} \right) }
  {\int_{\theta_0}^{\pi-\theta_0} d\theta_{l}
  \left( \frac{d\sigma^{+}}{d\theta_l}+\frac{d\sigma^{-}}{d\theta_l} \right) }.
\eqno{(4.1)}
$$
The charged leptonic forward-backward asymmetry with a charged lepton angle
cutoff $\theta_0$ is defined in the form as:
$$
{\cal A}_{FB}(\theta_0)=\frac{ \int_{\theta_0}^{\frac{\pi}{2}} d\theta_{l}
  \left( \frac{d\sigma^{+}}{d\theta_l}-\frac{d\sigma^{-}}{d\theta_l} \right) -
  \int_{\frac{\pi}{2}}^{\pi-\theta_0} d\theta_{l}
  \left( \frac{d\sigma^{+}}{d\theta_l}-\frac{d\sigma^{-}}{d\theta_l} \right) }
  { \int_{\theta_0}^{\pi-\theta_0} d\theta_{l}
  \left( \frac{d\sigma^{+}}{d\theta_l}+\frac{d\sigma^{-}}{d\theta_l} \right) }.
\eqno{(4.2)}
$$
These two types of angular asymmetries can be the measures of CP violation
in the process $e^{+} e^{-} \rightarrow t \bar{t}$ in unpolarized case.
The relevant expressions of $A_{ch}$ and $A_{FB}$ for this process can
be found in Ref.\cite{Poul}.

We use the input parameters as described in the above section and
take a leptonic angle-cutoff $\theta_{0}=30^{\circ}$ in numerical
calculations. The charged leptonic charge asymmetry $A_{ch}$ and
the forward-backward asymmetry $A_{FB}$ versus the center-of-mass
energy of $e^{+}~e^{-}$ system are depicted in Fig.4(1). In this
figure, the absolute value of the asymmetry $A_{ch}$, in general,
is smaller than that of $A_{FB}$ when the same parameters have
been taken. Actually, if the charged leptonic angle-cutoff
$\theta_{0}$ is changed smaller, the value of the asymmetry
$A_{ch}$ will decrease greatly, e.g. when the cutoff is changed
into $9^{\circ}$, the asymmetry $A_{ch}$ will decrease by an order
of magnitude comparing with that for the former choice. From this
figure we can see that the CP violation asymmetries have threshold
effects in the energy region near $2~ m_{t}$ clearly. Due to the
phase space effects, there is a small spike on each of the curves
in Fig.4(1). Our numerical calculation shows that when the leptonic
cutoff-angle is changed to a smaller value from $30^\circ$, the spikes
are shifted to the left side until it vanishes. The dependences of the
two types of the CP-violation asymmetries on the masses of Higgs bosons
$h^0$, $H^{\pm}$ and $A^0$ are plotted in Fig.4(2) and Fig.4(3),
respectively. The asymmetries depend on the masses of the CP-odd
neutral Higgs boson and the CP-even neutral Higgs boson $A^0$ and
$m_{h^{0}}$ precisely, but they are not sensitive to the mass of charged
Higgs boson $H^{\pm}$. Again, in Figs.4(2,3) we can see the effects
on the curves due to the resonances at the positions of
$m_{H^{\pm}}$ and $m_{h^0}$ as well as $m_{H^{\pm}}
\sim m_t+m_b \sim 180$ GeV and $m_{h^0} \sim 400$ GeV respectively.
The two asymmetries attributed to the electric and weak dipoles
versus the CP-violation phase angles $\theta_{1}(\theta^{U}_{33})$ and
$\theta_{2}(\theta^{D}_{33})$ are plotted in Fig.4(4) and Fig.4(5),
respectively. In each of these two figures, the calculated results are
obtained in each case by assuming one of the phase angles being zero
and with the same input parameters as those being taken before.
They show again that the CP-violation effects
mainly come from the phase angle $\theta_1$ and have maximal
absolute values when $\theta_{1}$ has the values about $125^\circ$ or
$235^\circ$. The quantitative contributions to the asymmetries due to
the CP-violating phase angle $\theta_2$ are smaller than those due to
the phase angle $\theta_1$.

There are alternative choices for the CP-odd observables to
observe CP violation, of which some can be better, but some not.
Moreover, to distinguish different CP violation sources,
various CP-odd observables are needed in general.
For the purpose and to try one more observable, let us
introduce a new CP asymmetric parameter by means
of the vectors $\vec{s}$ and $\vec{\omega}$, i.e., the spin and
the polarization vectors of the top quark respectively.
The two vectors are three dimensional essentially.
The vector $\vec{s}$ is normalized by definition,
and the vector $\vec{\omega}_{t}$ is
defined as\cite{Tsai}
$$
\omega_{t,i}=\frac
                {N_{t}(\vec{s}=\hat{e}_{i})-
                 N_{t}(\vec{s}=-\hat{e}_{i}) }
                {N_{t}(\vec{s}=\hat{e}_{i})+
                 N_{t}(\vec{s}=-\hat{e}_{i}) }.
\eqno{(4.3)}
$$
where $N_{t}(\vec{s}=\hat{e}_{i})$ means the event number of the
top quark with a spin vector as $\vec{s}=\hat{e}_{i}$.
Here $\hat{e}_{i}$ (i=1, 2, 3) are the normalized
basis vectors of a Cartesian coordinate frame. Note that
in the rest frame of the top quark, the four-dimensional
spin vector $s_\mu$ of the top quark becomes $(0,\vec{s})$.
To calculate the polarization vector
$\omega_{t,i}$ in the CMS of electron and positron,
we shall introduce the coordinate system (x', y', z'), where z'-axis is along
the outgoing direction of top quark and y'-axis is perpendicular to the
production plane of top quark pair.
The commonly used coordinate system (x, y, z) in the CMS of
initial states, is defined as:
z-axis is along the incoming direction of $e^-$ and y-axis is
perpendicular to the production plane of top pair.
The angle between the axes z and z' is just the angle $\theta$ between
the incoming electron and the outgoing top quark. Since we do not
observe both polarizations of top and anti-top quarks
in the meantime, thus to focus the polarization of the top quark,
we sum up the spin of anti-top quark when calculating polarization
vector $\vec{\omega}_{t}$. In this way, for the top quark spin
$\vec{s}=\hat{e}_{z'}$, the corresponding four-dimension spin vector
in (x', y', z') system becomes as\cite{Tsai}
$$
(s)_{\mu}=(\beta \gamma,0,0,\gamma),
\eqno{(4.4)}
$$
where $\gamma=\frac{\sqrt{s}}{2 m_{t}}$ and $\beta=\sqrt{1-\gamma^{-2}}$.
For the top quark spin vectors $\vec{s}=\hat{e}_{x'},\hat{e}_{y'}$, the
corresponding four-dimension spin vectors are
$$
(s)_{\mu}=(0,1,0,0)~~~and~~~
(s)_{\mu}=(0,0,1,0),
\eqno{(4.5)}
$$
respectively.

Analogously, the vectors $\vec{\omega}_{\bar{t}}$ for the top anti-quark
can be defined too. Now let us introduce the so-called
spin-momentum correlation-parameter for the top quark outgoing
in a specific direction,
$$
\xi_{CP}=(\hat{p}_{e^{-}} \times \hat{p}_{t}) \cdot
         (\vec{\omega}_{t}-\vec{\omega}_{\bar{t}}),
\eqno{(4.6)}
$$
and it is easy to see that the correlation parameter
is CP-odd. Here $\hat{p}_{e^{-}}$ and $\hat{p}_{t}$ are unit
vectors of the three-dimensional momenta of the
electron and the top quark, and
$\vec{\omega}_{t}$ and $\vec{\omega}_{\bar{t}}$ are
the polarization vectors of the top quark and the
top anti-quark in the (x, y, z) system.
We know that if CP conservation is hold, the
polarization vectors $\vec{\omega}_{t}$ and $\vec{\omega}_{\bar{t}}$ of the
top quark and the top anti-quark cannot have nonzero components
perpendicular to the production plane
$i.e., \omega_{t,y}=\omega_{\bar{t},y}=0$, and consequently the parameter
$\xi_{CP}$ must be zero. If CP violates in the process, only the components
of the polarization vectors $\vec{\omega}_{t}$ and $\vec{\omega}_{\bar{t}}$
perpendicular to the production plane($\omega_{y}$)
contribute to the asymmetry parameter $\xi_{CP}$. Thus to
calculate the asymmetry observable $\xi_{CP}$, we have to
compute the polarization vectors of the top quark and the top anti-quark
in y-component only. Namely the observable $\xi_{CP}$
can be measured by the possible asymmetry in the polarization vectors of
$t$ and $\bar{t}$ perpendicular to the production plane.

Taking a cutoff for the top quark angle $\theta_{cut}=9^\circ$,
we compute the spin-momentum correlation parameter $\xi_{CP}$
numerically by the above definitions and
plot its average value $<\xi_{CP}>$
as a function of $\sqrt{s}$ in Fig.5(1).
The threshold influence at the energy region near $\sqrt{s} \sim 2~m_t$
and smooth variation (approaches to $4 \times 10^{-4}$ with the increasing
of $\sqrt{s}$) of the average value of the asymmetry parameter $<\xi_{CP}>$,
are shown clearly in the figure.
The dependences of the average value of the CP-violation parameter,
$<\xi_{CP}>$, on the masses of Higgs bosons are represented in Fig.5(2).
There are similar dependences on the Higgs boson masses
as those illustrated in the previous cases
for the electric and weak dipoles of the top quark and the other
observables. We can also see that the
resonance effects are manifested on the curves
for the dependences on the masses $m_{h^0}$ and $m_{H^{\pm}}$,
e.g. when $m_{h^0}=400~GeV$ and $m_{H^{\pm}}=m_{t}+m_{b}$,
respectively. The parameter $<\xi_{CP}>$ versus CP-violation angles
$\theta_1$ and $\theta_2$ are depicted in Fig.5(3) respectively, by assuming
the CP-violation angles $\theta_2$ and $\theta_1$ to be zero in turn.
Thus it is seen clearly that the
CP-violation here takes source mainly from the phase angle $\theta_1$ and
has the maximal absolute value when $\theta_{1}$ has the values about
$125^\circ$ and $235^\circ$, whereas the contribution
to the asymmetry parameter due to the CP-violating phase angle $\theta_2$
is much smaller than that due to the
phase angle $\theta_1$.

\vskip 5mm
\noindent{\large\bf 5. Summary}

We have calculated the contributions to the electric and weak dipoles of
the top quark in the frame of the THDM-III by introducing complex
phase parameters $\xi^{U}$ and $\xi^{D}$. The complete analytical
expressions of the dipole form factors of top quark are presented.
Some observables which are sensitive to CP violation
are discussed. The form factors for the electric and weak dipoles, the
CP-violating observables $A_{ch}$, $A_{FB}$ and $<\xi_{CP}>$ all
depend on the masses of Higgs bosons, the CP-violating phase angles
$\theta_{1}$ and $\theta_{2}$ and the CMS energy in electron-positron
system as well.

We have found that the electric and
weak dipole form factors and other observables are enhanced in the
threshold region of top pair production. The CP-violation
asymmetries are mostly related to the neutral Higgs bosons $A^{0}$ and $h^0$
masses and the phase angle $\theta_{1}$. The numerical results show that they
are in general of order of $10^{-4}$, and the electric, weak dipoles
and the spin-momentum asymmetry $<\xi_{CP}>$ can reach about $10^{-3}$
quantitatively. In the sense, being a directly measurable observable,
$<\xi_{CP}>$ may be a better one than the others.

In summary, the values for the CP-odd observables considered in the paper
may fall into the ability for the underconsidering
colliders such as NLC, which can reach to a high enough energy and
luminosity i.e., the asymmetries may become measurable for the powerful
colliders. Therefore we would like to conclude that
the top quark pair production at a very high energy $e^+e^-$ collider
may serve as a process for probing and hopefully
investigating the CP violation sources in the THDM-III by
measuring various CP-odd observables.

\vspace{4mm}
\noindent{\large\bf Acknowledgement:}
These work was supported in part by the National Natural Science
Foundation of China (project numbers: 19675033 and 19677102)
and the Grant of Chinese Academy of Science. The authors
would like to thank Dr. J. P. Ma for useful comments.

\vspace{8mm}
\noindent{\large\bf Appendix}

The electric dipole form factor $d_{\gamma}$, taking the source of
the THDM-III, can be presented explicitly as:
$$
\begin{array}{lll}
d_{\gamma} &=&\frac{ i m_{t}^2 g^2}{96 m_{W}^2 \pi^2}(3 (e^{(i \theta^{D} +
i \theta^{U})}- e^{(-i \theta^{D} - i \theta^{U})}) m_{b}^2 C_{0}^{1}\\
&+&(e^{(2 i \theta^{U})}-e^{(-2 i \theta^{U})}) m_{t}^2 C_{11}^{2}
-(e^{(i \theta^{D} + i \theta^{U})}-e^{(-i \theta^{D} -
i \theta^{U})}) m_{b}^2 C_{11}^{3} \\
&+&(m_{t}^2-m_{b}^2) C_{11}^{3}
-(e^{(2 i \theta^{U})}-e^{(-2 i \theta^{U})}) m_{t}^2
\cos^2\alpha C_{11}^{4} \\
&-&3 (m_{t}^2-m_{b}^2) C_{11}^{1}
+3 (e^{(i \theta^{D} + i \theta^{U})}-
e^{(-i \theta^{D} - i \theta^{U})}) m_{b}^2 C_{11}^{1} \\
&-&2 (m_{t}^2-m_{b}^2) C_{12}^{3}
+6 (m_{t}^2-m_{b}^2) C_{12}^{1} \\
&+&(m_{t}^2-m_{b}^2) C_{21}^{3}
-3 (m_{t}^2-m_{b}^2) C_{21}^{1} \\
&-&2 (m_{t}^2-m_{b}^2) C_{23}^{3}
+6 (m_{t}^2-m_{b}^2) C_{23}^{1} \\
&-&2 (e^{(i \theta^{U})}-e^{(-i \theta^{U})}) m_{t}^2 \cos\alpha
C_{11}^{4} \sin\alpha \\
&+&2 (e^{(i \theta^{U})}-e^{(-i \theta^{U})}) m_{t}^2 \cos\alpha
C_{11}^{5} \sin\alpha \\
&-&(e^{(2 i \theta^{U})}-e^{(-2 i \theta^{U})}) m_{t}^2
C_{11}^{5} \sin^2\alpha)
\hskip 15mm (A.1)
\end{array}
$$

The weak dipole form factor $d_{Z}$ due to the THDM-III
can be presented explicitly as follows:
$$
\begin{array}{lll}
d_{z}&=&\frac{i m_{t}^2 g^2}{(384 m_W^2\pi^2)\sin{\theta_{w}}
  \cos{\theta_{w}}}( 6(e^{(i\theta^{D} + i\theta^{U})}-
  e^{(-i\theta^{D} - i\theta^{U})})m_{b}^2
   \cos{2\theta_{w}} C_{0}^{1} \\
&+&3(e^{(2i\theta^{U})}-1)m_{t}^2 C_{11}^{2}
+6(m_{t}^2-e^{(i\theta^{D} + i\theta^{U})}m_{b}^2) C_{11}^{3} \\
&-&3(e^{(2i\theta^{U})}+1)m_{t}^2\cos^2{\alpha}
        C_{11}^{4}  \\
&-&6m_{t}^2\cos^2{\alpha} C_{11}^{5} + 6(e^{(i\theta^{D} +
i\theta^{U})}-e^{(-i\theta^{D} - i\theta^{U})})m_{b}^2\cos{2\theta_{w}}
          C_{11}^{1} \\
&-&6(m_{t}^2-m_{b}^2)\cos{2\theta_{w}} C_{11}^{1}
+3(2-e^{(-2i\theta^{U})}-e^{(2i\theta^{U})})m_{t}^2 C_{12}^{2}    \\
&+&6(e^{(-i\theta^{D} - i\theta^{U})}+e^{(i\theta^{D}
+ i\theta^{U})})m_{b}^2 C_{12}^{3}
-12m_{t}^2 C_{12}^{3}   \\
&+&3(2+e^{(-2i\theta^{U})}+e^{(2i\theta^{U})})m_{t}^2\cos^2{\alpha}
C_{12}^{4}
+12m_{t}^2\cos^2{\alpha} C_{12}^{5}  \\
&+&12(m_{t}^2-m_{b}^2)\cos{2\theta_{w}} C_{12}^{1}
-3m_{t}^2 (C_{21}^{2}-2 C_{21}^{3})
-3m_{t}^2\cos^2{\alpha} (C_{21}^{4}+C_{21}^{5}) \\
&-&6(m_{t}^2-m_{b}^2)\cos{2\theta_{w}} C_{21}^{1}
+ 6m_{t}^2 (C_{23}^{2}-2 C_{23}^{3})
+6m_{t}^2\cos^2{\alpha} (C_{23}^{4}+C_{23}^{5})  \\
&+&12(m_{t}^2-m_{b}^2)\cos{2\theta_{w}} C_{23}^{1}
-3(e^{(-i\theta^{U})}+3e^{(i\theta^{U})})m_{t}^2\cos{\alpha} C_{11}^{4}
         \sin{\alpha}  \\
&+&3(e^{(-i\theta^{U})}+3e^{(i\theta^{U})})m_{t}^2\cos{\alpha}
   C_{11}^{5} \sin{\alpha} \\
&+&3(e^{(-i\theta^{U})}+e^{(i\theta^{U})})m_{t}^2\cos{\alpha}
   (4 C_{12}^{4}-4 C_{12}^{5}-C_{21}^{4}+C_{21}^{5}+2 C_{23}^{4}-2 C_{23}^{5})
    \sin{\alpha} \\
&+&3m_{t}^2 (-2 C_{11}^{4}-(1+e^{(2i\theta^{U})})C_{11}^{5}
   +4 C_{12}^{4}+(2+e^{(-2i\theta^{U})}
+e^{(2i\theta^{U})}) C_{12}^{5}-C_{21}^{4}\\
&-&C_{21}^{5}+2 C_{23}^{4}+2 C_{23}^{5})\sin^2{\alpha}
+4(e^{(-2i\theta^{U})}-e^{(2i\theta^{U})})m_{t}^2
C_{11}^{2}\sin^{2}{\theta_{w}} \\
&+&4(e^{(i\theta^{D} + i\theta^{U})}-e^{(-i\theta^{D}
-i\theta^{U})})m_{b}^2C_{11}^{3}\sin^{2}{\theta_{w}}  \\
&-&4(m_{t}^2-m_{b}^2)C_{11}^{3}\sin^{2}{\theta_{w}}
+4(e^{(2i\theta^{U})}-e^{(-2i\theta^{U})})
m_{t}^2\cos^2{\alpha}C_{11}^{4}\sin^{2}{\theta_{w}}   \\
&+&4(m_{t}^2-m_{b}^2)(2 C_{12}^{3}-C_{21}^{3}+
2C_{23}^{3})\sin^{2}{\theta_{w}} \\
&+&8(e^{(i\theta^{U})}-e^{(-i\theta^{U})})m_{t}^2\cos{\alpha}
   (C_{11}^{4}-C_{11}^{5})\sin{\alpha}\sin^{2}{\theta_{w}}  \\
&+&4(e^{(2i\theta^{U})}-e^{(-2i\theta^{U})})m_{t}^2
   C_{11}^{5}\sin^2{\alpha}\sin^{2}{\theta_{w}} \\
&+& 2(C_{11}^{6} - C_{12}^{6} - C_{12}^{7})(e^{-i\theta^{U}} -
    e^{i\theta^{U}})m_{Z}^2 (4\cos{2\theta_W} - 1)\sin{2\alpha} )
\hskip 25mm (A.2)
\end{array}
$$
Here the following notations are adopted:
$$
\begin{array}{lll}
C_{i,ij}^{1}&=&C_{i,ij}[p_{2}, -p_{1} - p_{2},
m_{b}, m_{H^{\pm}}, m_{H^{\pm}}]\\
C_{i,ij}^{2}&=&C_{i,ij}[p_{1}, -p_{1} - p_{2}, m_{A^{0}}, m_{t}, m_{t}]   \\
C_{i,ij}^{3}&=&C_{i,ij}[p_{1}, -p_{1} - p_{2}, m_{H^{\pm}}, m_{b}, m_{b}] \\
C_{i,ij}^{4}&=&C_{i,ij}[p_{1}, -p_{1} - p_{2}, m_{h^{0}}, m_{t}, m_{t}]   \\
C_{i,ij}^{5}&=&C_{i,ij}[p_{1}, -p_{1} - p_{2}, m_{H^{0}}, m_{t}, m_{t}]   \\
C_{i,ij}^{6}&=&C_{i,ij}[-p_{1}, p_{1} + p_{2}, m_{t}, m_{Z}, m_{H}]       \\
C_{i,ij}^{7}&=&C_{i,ij}[-p_{1}, p_{1} + p_{2}, m_{t}, m_{h}, m_{Z}]
\hskip 15mm (A.3)
\end{array}
$$
In above expressions we adopted the definition of the three-point
integral functions in Ref.\cite{abcd} and the references therein.

\vskip 20mm


\vskip 20mm
\noindent{\Large\bf Figure captions}
\vskip 5mm

\noindent
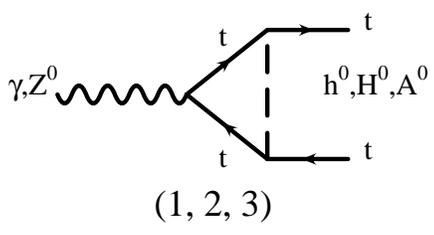
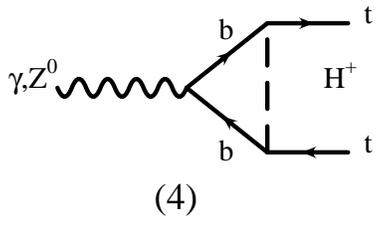
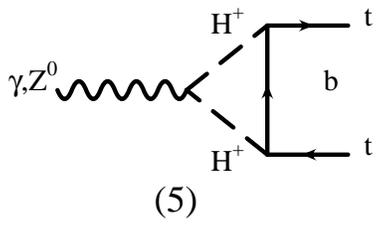
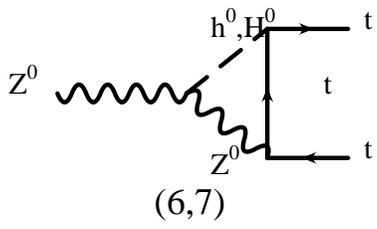
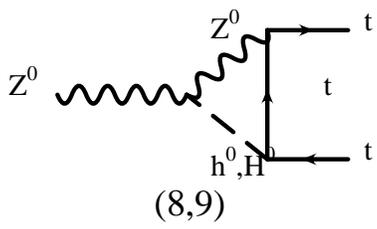
{\bf Fig.1} The Feynman diagrams which contribute to electric and weak
dipoles of top quark.
\vskip 3mm

\noindent
{\bf Fig.2} The Feynman rules of the relevant vertices involving
complex coupling constants.
\vskip 3mm

\noindent
{\bf Fig.3(1)} The electric and weak dipoles $d_{\gamma}$ and $d_{Z}$ as
functions of the c.m.s. energy $\sqrt{s}$. The full line is for
$Re[d_{\gamma}]$. The dashed line is for $Im[d_{\gamma}]$. The dotted
line is for $Re[d_{Z}]$. The dash-dotted line is for $Im[d_{Z}]$. In the
figure we take $m_{h^{0}}=100~GeV$, $m_{H^{0}}=200~GeV$, $m_{H^{\pm}}=150~GeV$,
$m_A=250~GeV$, $\alpha=\frac{\pi}{6}$, $\theta_1=\frac{\pi}{4}$
and $\theta_2=\frac{\pi}{2}$.
\vskip 3mm

\noindent
{\bf Fig.3(2)} The electric and weak dipoles $d_{\gamma}$ and $d_{Z}$ as
functions of the mass of $h^0$. The full line is for
$Re[d_{\gamma}]$. The dashed line is for $Im[d_{\gamma}]$. The dotted
line is for $Re[d_{Z}]$. The dash-dotted line is for $Im[d_{Z}]$. In the
figure we take $\sqrt{s}=500~GeV$, $m_{H^{0}}=200~GeV$, $m_{H^{\pm}}=150~GeV$,
$m_A=250~GeV$, $\alpha=\frac{\pi}{6}$, $\theta_1=\frac{\pi}{4}$
and $\theta_2=\frac{\pi}{2}$.
\vskip 3mm

\noindent
{\bf Fig.3(3)} The electric and weak dipoles $d_{\gamma}$ and $d_{Z}$ as
functions of the mass of $H^{\pm}$. The full line is for
$Re[d_{\gamma}]$. The dashed line is for $Im[d_{\gamma}]$. The dotted
line is for $Re[d_{Z}]$. The dash-dotted line is for $Im[d_{Z}]$. In the
figure we take $\sqrt{s}=500~GeV$, $m_{h^{0}}=100~GeV$, $m_{H^{0}}=200~GeV$,
$m_A=250~GeV$, $\alpha=\frac{\pi}{6}$, $\theta_1=\frac{\pi}{4}$
and $\theta_2=\frac{\pi}{2}$.
\vskip 3mm

\noindent
{\bf Fig.3(4)} The electric and weak dipoles $d_{\gamma}$ and $d_{Z}$ as
functions of the mass of $A^0$. The full line is for
$Re[d_{\gamma}]$. The dashed line is for $Im[d_{\gamma}]$. The dotted
line is for $Re[d_{Z}]$. The dash-dotted line is for $Im[d_{Z}]$. In the
figure we take $\sqrt{s}=500~GeV$, $m_{h^{0}}=100~GeV$, $m_{H^{0}}=200~GeV$,
$m_{H^{\pm}}=150~GeV$, $\alpha=\frac{\pi}{6}$, $\theta_1=\frac{\pi}{4}$
and $\theta_2=\frac{\pi}{2}$.
\vskip 3mm

\noindent
{\bf Fig.3(5)} The electric and weak dipoles $d_{\gamma}$ and $d_{Z}$ as
functions of the phase angle $\theta_{1}$. The full line is for
$Re[d_{\gamma}]$. The dashed line is for $Im[d_{\gamma}]$. The dotted
line is for $Re[d_{Z}]$. The dash-dotted line is for $Im[d_{Z}]$. In the
figure we take $\sqrt{s}=500~GeV$, $m_{h^{0}}=100~GeV$, $m_{H^{0}}=200~GeV$,
$m_{H^{\pm}}=150~GeV$, $m_A=250~GeV$, $\alpha=\frac{\pi}{6}$ and $\theta_2=0$.
\vskip 3mm

\noindent
{\bf Fig.3(6)} The electric and weak dipoles $d_{\gamma}$ and $d_{Z}$ as
functions of the phase angle $\theta_{2}$. The full line is for
$Re[d_{\gamma}]$. The dashed line is for $Im[d_{\gamma}]$. The dotted
line is for $Re[d_{Z}]$. The dash-dotted line is for $Im[d_{Z}]$. In the
figure we take $\sqrt{s}=500~GeV$, $m_{h^{0}}=100~GeV$, $m_{H^{0}}=200~GeV$,
$m_{H^{\pm}}=150~GeV$, $m_A=250~GeV$, $\alpha=\frac{\pi}{6}$ and
$\theta_1=0$.
\vskip 3mm

\noindent
{\bf Fig.4(1)} The lepton charge distribution asymmetry $A_{ch}$
and the forward-backward asymmetry $A_{FB}$ as functions of the c.m.s.
energy $\sqrt{s}$
for the charged lepton (the product of the sequence decays $t\to b+W$ and
$W\to l+\nu$). The full line is for $A_{ch} \times 10$. The dashed line
is for $A_{FB}$. In the figure we take $m_{h^{0}}=100~GeV$, $m_{H^{0}}=200~GeV$,
$m_{H^{\pm}}=150~GeV$, $m_A=250~GeV$, $\alpha=\frac{\pi}{6}$,
$\theta_1=\frac{\pi}{4}$, $\theta_2=\frac{\pi}{2}$, and
$\theta_{0}=30^{\circ}$.
\vskip 3mm

\noindent
{\bf Fig.4(2)} The forward-backward asymmetry $A_{FB}$ as a function
of the masses of Higgs bosons $h^0$, $H^0$ and $H_{\pm}$. The full line
is for $m_{h^0}$ with $m_{H^{\pm}}=150~GeV$ and $m_A=250~GeV$. The dashed
line is for $m_{H^0}$ with $m_{h^{0}}=100~GeV$ and $m_A=250~GeV$. The dotted
line is for $m_{A^0}$ with $m_{h^{0}}=100~GeV$ and $m_{H^{\pm}}=150~GeV$.
In the figure we take $\sqrt{s}=500~GeV$, $m_H^{0}=200~GeV$,
$\alpha=\frac{\pi}{6}$, $\theta_1=\frac{\pi}{4}$,
$\theta_2=\frac{\pi}{2}$, and $\theta_{0}=30^{\circ}$.
\vskip 3mm

\noindent
{\bf Fig.4(3)} The lepton charge asymmetry $A_{ch}$ as a function
of the masses of Higgs bosons $h^0$, $H^0$ and $H_{\pm}$. The full line
is for $m_{h^0}$ with $m_H^{\pm}=150~GeV$ and $m_A=250~GeV$. The dashed
line is for $m_{H^0}$ with $m_{h^{0}}=100~GeV$ and $m_A=250~GeV$. The dotted
line is for $m_{A^0}$ with $m_{h^{0}}=100~GeV$ and $m_{H^{\pm}}=150~GeV$.
In the figure we take $\sqrt{s}=500~GeV$, $m_H^{0}=200~GeV$,
$\alpha=\frac{\pi}{6}$, $\theta_1=\frac{\pi}{4}$,
$\theta_2=\frac{\pi}{2}$, and $\theta_{0}=30^{\circ}$.
\vskip 3mm

\noindent
{\bf Fig.4(4)} The lepton charge distribution asymmetry $A_{ch}$
and the forward-backward asymmetry $A_{FB}$ as functions of the phase
angle $\theta_1$. The full line is for $A_{ch} \times 10$. The dashed line
is for $A_{FB}$. In the figure we take $m_{h^{0}}=100~GeV$, $m_{H^{0}}=200~GeV$,
$m_{H^{\pm}}=150~GeV$, $m_A=250~GeV$, $\alpha=\frac{\pi}{6}$,
$\theta_1=\frac{\pi}{4}$, $\theta_2=\frac{\pi}{2}$ and
$\theta_{0}=30^{\circ}$.
\vskip 3mm

\noindent
{\bf Fig.4(5)} The lepton charge distribution asymmetry $A_{ch}$
and the forward-backward asymmetry $A_{FB}$ as functions of the phase
angle $\theta_2$. The full line is for $A_{ch} \times 10$. The dashed line
is for $A_{FB}$. In the figure we take $m_{h^{0}}=100~GeV$, $m_{H^{0}}=200~GeV$,
$m_{H^{\pm}}=150~GeV$, $m_A=250~GeV$, $\alpha=\frac{\pi}{6}$,
$\theta_1=\frac{\pi}{4}$, $\theta_2=\frac{\pi}{2}$ and
$\theta_{0}=30^{\circ}$.
\vskip 3mm

\noindent
{\bf Fig.5(1)} The average value of the spin-momentum correlation parameter
$\xi_{CP}$ as a function of the c.m.s. energy $\sqrt{s}$. In the figure
we take $m_{h^{0}}=100~GeV$, $m_{H^{0}}=200~GeV$, $m_{H^{\pm}}=150~GeV$,
$m_A=250~GeV$, $\alpha=\frac{\pi}{6}$, $\theta_1=\frac{\pi}{4}$,
$\theta_2=\frac{\pi}{2}$, and $\theta_{cut}=9^{\circ}$.
\vskip 3mm

\noindent
{\bf Fig.5(2)} The average value of the spin-momentum correlation parameter
$\xi_{CP}$ as a function of the masses of Higgs bosons. The full line is
for $m_{h^{0}}$ with $m_{H^{\pm}}=150~GeV$ and $m_A=250~GeV$. The dashed line
is for $m_{H^{\pm}}$ with $m_{h^{0}}=100~GeV$ and $m_A=250~GeV$. The dotted
line is for $m_{A}$ with $m_{h^{0}}=100~GeV$ and $m_{H^{\pm}}=150~GeV$.
In the figure we take $m_H^{0}=200~GeV$, $\alpha=\frac{\pi}{6}$,
$\theta_1=\frac{\pi}{4}$, $\theta_2=\frac{\pi}{2}$ and
$\theta_{cut}=9^{\circ}$.
\vskip 3mm

\noindent
{\bf Fig.5(3)} The average value of the spin-momentum correlation parameter
$\xi_{CP}$ as function of the phase angles $\theta_1$ and $\theta_2$.
The full line is for $\theta_1$ with $\theta_2=0$. The dashed line
is for $\theta_2$ with $\theta_1=0$. In the figure we take $\sqrt{s}=500~GeV$,
$m_{h^{0}}=100~GeV$, $m_{H^{0}}=200~GeV$, $m_{H^{\pm}}=150~GeV$
$m_A=250~GeV$, $\alpha=\frac{\pi}{6}$ and $\theta_{cut}=9^{\circ}$.
\vskip 3mm
\noindent

\vskip 3mm

\begin{thebibliography}{s25}
\bibitem{CDF} CDF Collaboration, F. Abe et al., Phys. Rev. Lett.{\bf 74},
              2626(1995);
              D0 Collaboration, S. Abachi et al., $ibid.$ {\bf 74}, 2632(1995);
              P.C. Bhat, for the D0 collaboration, talk presented at the Wine
              and Cheese Seminar at Fermilab, February 1997.
\bibitem{Christ} J.H.Christenson, J.W. Cronin, V.L. Fitch and R. Turlay
              Phys. Rev. Lett.{\bf 13}, 138(1964).
\bibitem{Lee} T.D. Lee, Phys. Rev. {\bf D8}, 1226(1973); S. Weinberg,
              Phys. Rev. {\bf D42}, 860(1990).
\bibitem{Wu} L. Wolfenstein and Y.-L. Wu, Phys. Rev. Lett. {\bf 73}, 2809(1994).
\bibitem{Bern} W. Bernrouther, O. Nachtmann, P.Overmann and T. Schroder,
             Nucl. Phys. {\bf B388}, 53(1992); W. Bernrouther and O. Overmann,
             Z. Phys. {\bf C61}, 599(1994); W. Bernrouther and O. Nachtmann,
             Phys. Lett. {\bf B268}, 424(1991); W. Bernrouther
             and A. Brandenburg,
             Phys. Lett. {\bf B314}, 104(1993); Phys. Rev. {bf D49}, 4481(1994);
             A. Brandenburg, J.P. Ma, R. Munch and O. Nachtmann, Z. Phys.
             {bf C51}, 225(1991.
\bibitem{Christ1} E. Christova and M. Brandenburg, Phys. Lett. {\bf B315},
             338(1993); E. Christova and M. Fabbrichesi, Phys. Lett. {\bf B315},
             113(1993); D. Atwood and A. Soni, Phys. Rev. {\bf D45}, 2405(1992);
             A. Bartl, E. Christova and W. Majerotto, Nucl. Phys. {\bf B 460},
             235(1996).
\bibitem{Schmidt} C.R. Schmidt and M.E. Peskin, Phys. Rev. Lett. {\bf 69},
             410(1992); B. Grzadkowski and J. Gunion, Phys. Lett.
             {\bf B287}, 237(1992);
             C.R. Schmidt, Phys. Lett. {\bf B293}, 111(1992); E. Christova and
             M. Fabbrichesi, Phys. Lett. {\bf B320}, 299(1994); B. Grzadkowski
             W. Keung, Phys. Lett. {\bf B319}, 526(1993);
             B. Grzadkowski, Phys. Lett.
             {\bf B305}, 384(1993); B. Grzadkowski,
             Phys. Lett. {\bf B316}, 137(1993).
             A.Bartl, et. al, `Electroweak dipole moment form factors of the top
             quark in supersymmetry', hep-9709219.
\bibitem{Luke} M. Luke and M.J. Savage, Phys. Lett. {\bf B307}, 387(1993).
\bibitem{chengs} T. P. Cheng and M. Sher, Phys. Rev. {\bf D35}, 3484(1987);
             M. Sher and Y. Yuan, $ibid.$ {\bf 44}, 1461(1991).
\bibitem{Poul} P. Poulose and S.D. Rindani, Phys. Rev. {\bf D54}, 4326(1996).
\bibitem{Barn} Particle Data Group, R.M. Barnett, etal, Phys. Rev. {\bf D54},
             1(1996).
\bibitem{Tsai} Y.S. Tsai, Phys. Rev. {\bf D51}, 3172(1995).
\bibitem{abcd} Kniehl B. A., Phys. Rep. 240(1994)211 and references therein.
\end{thebibliography}
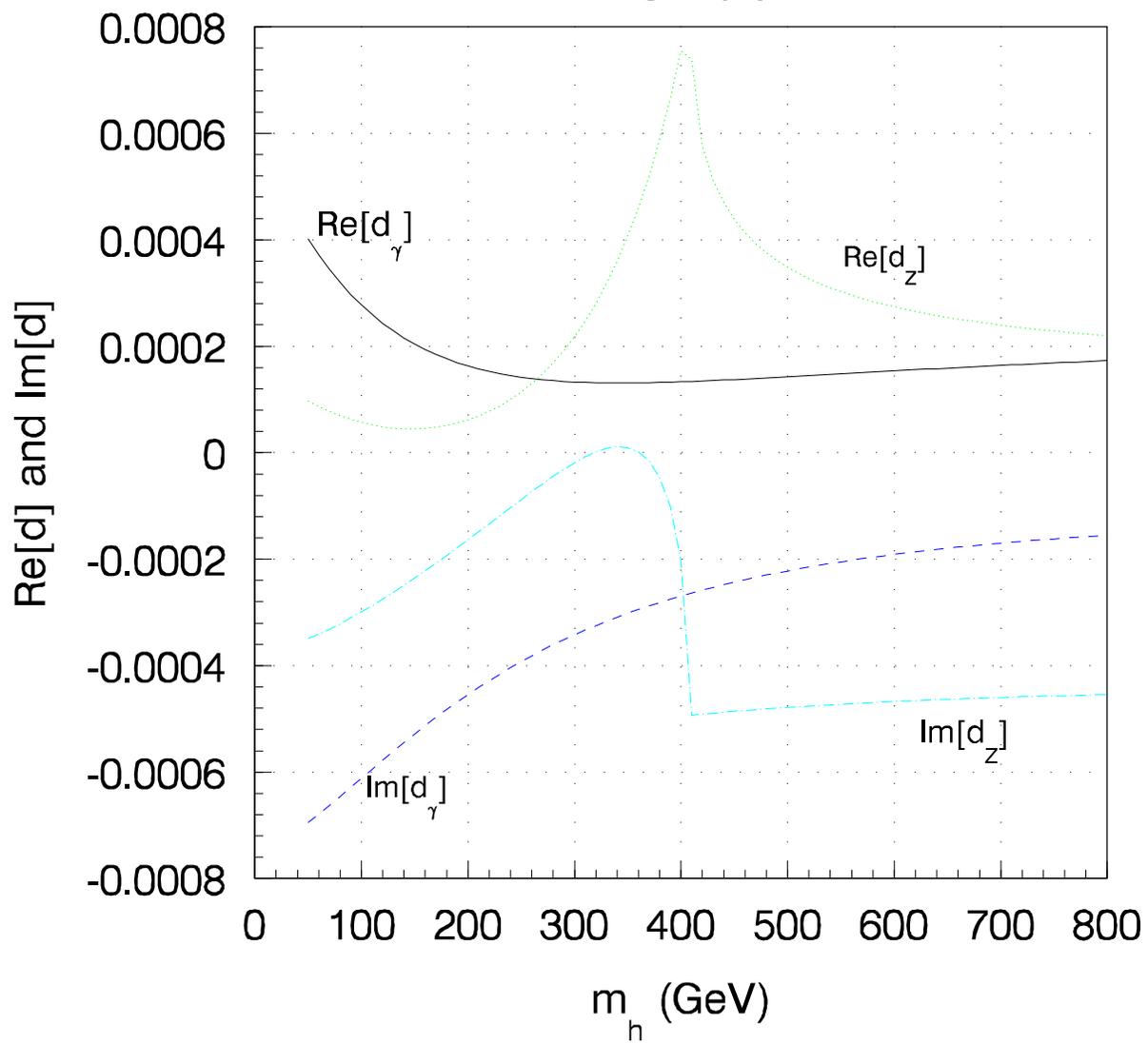
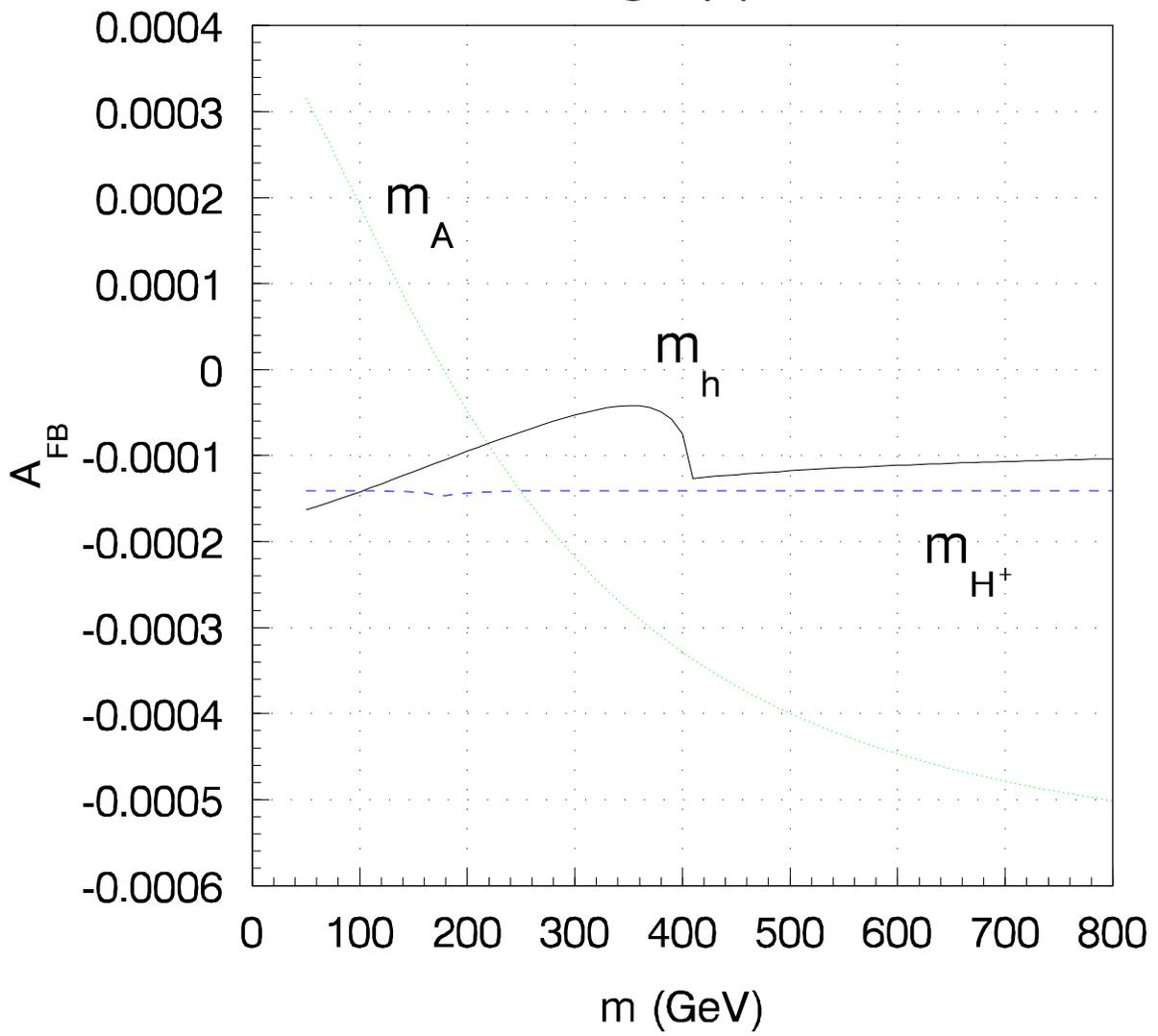
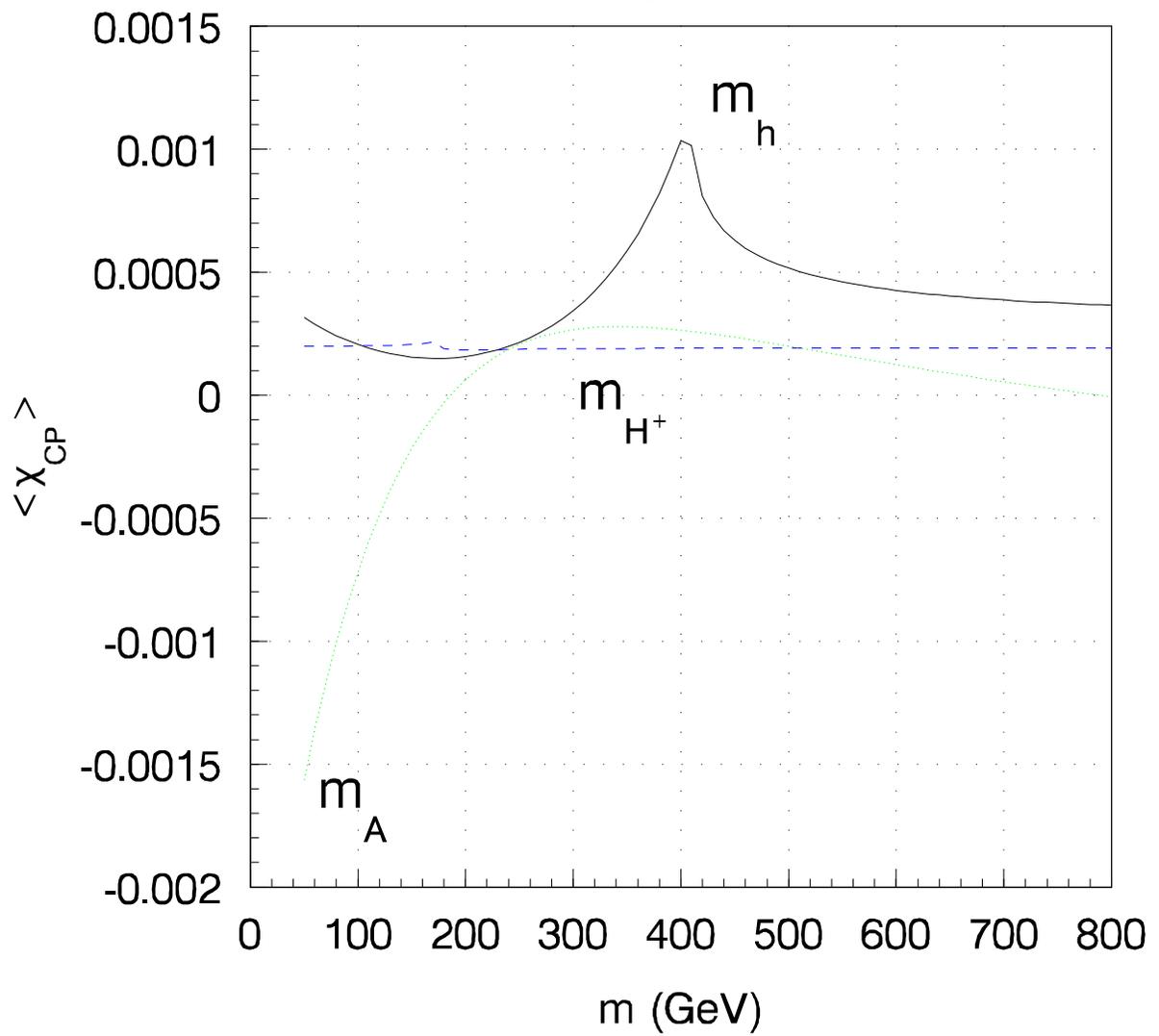
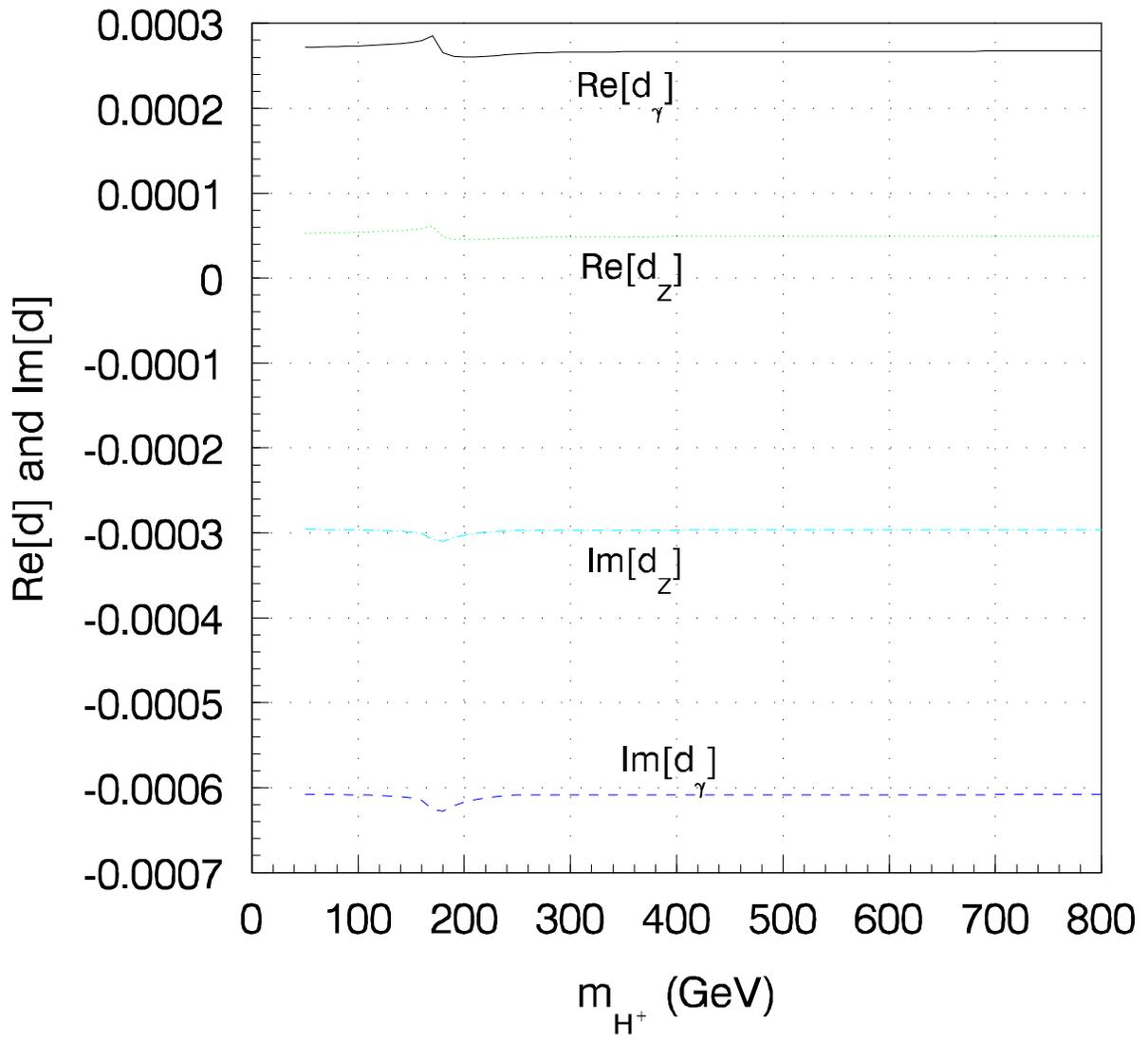
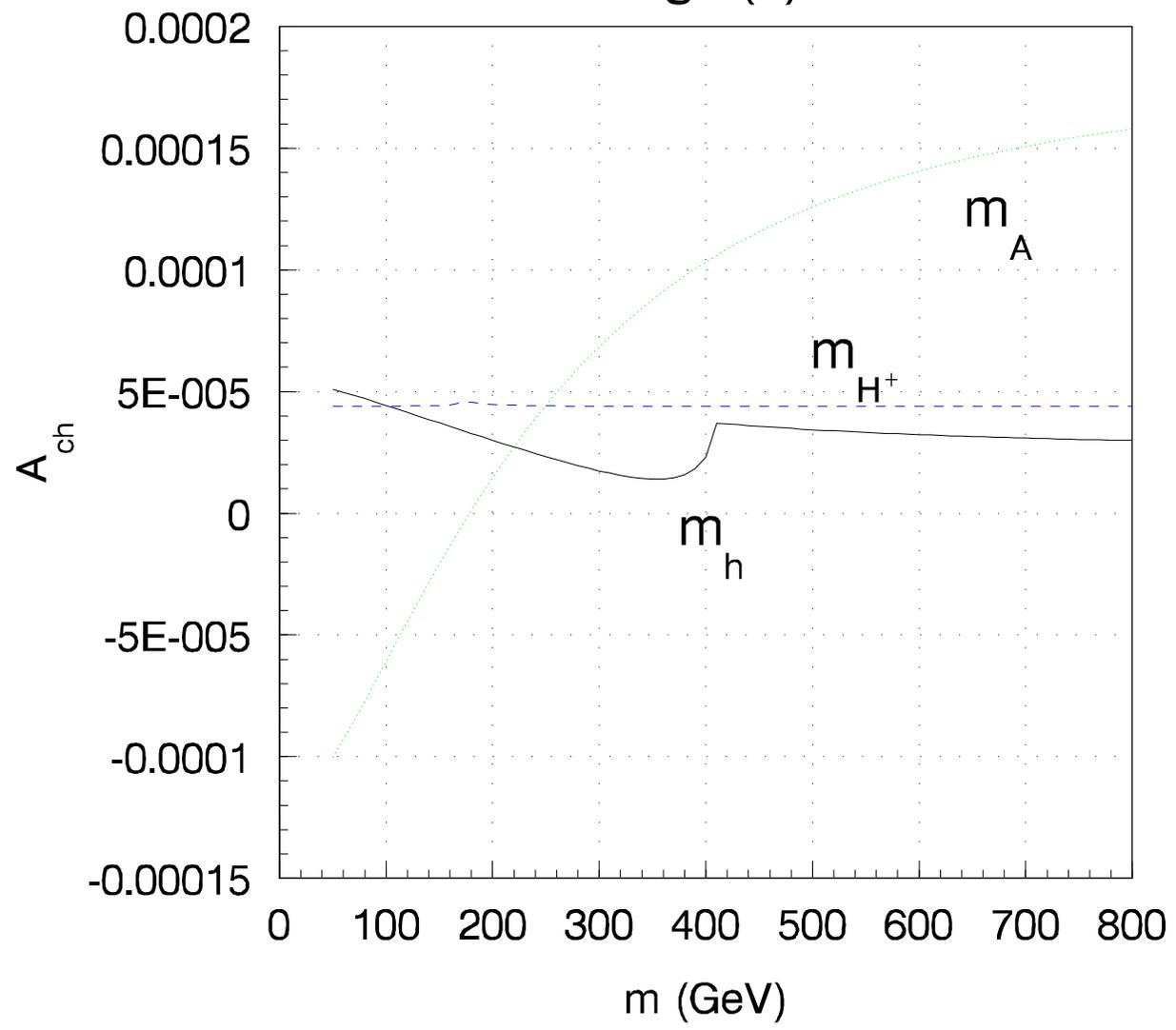
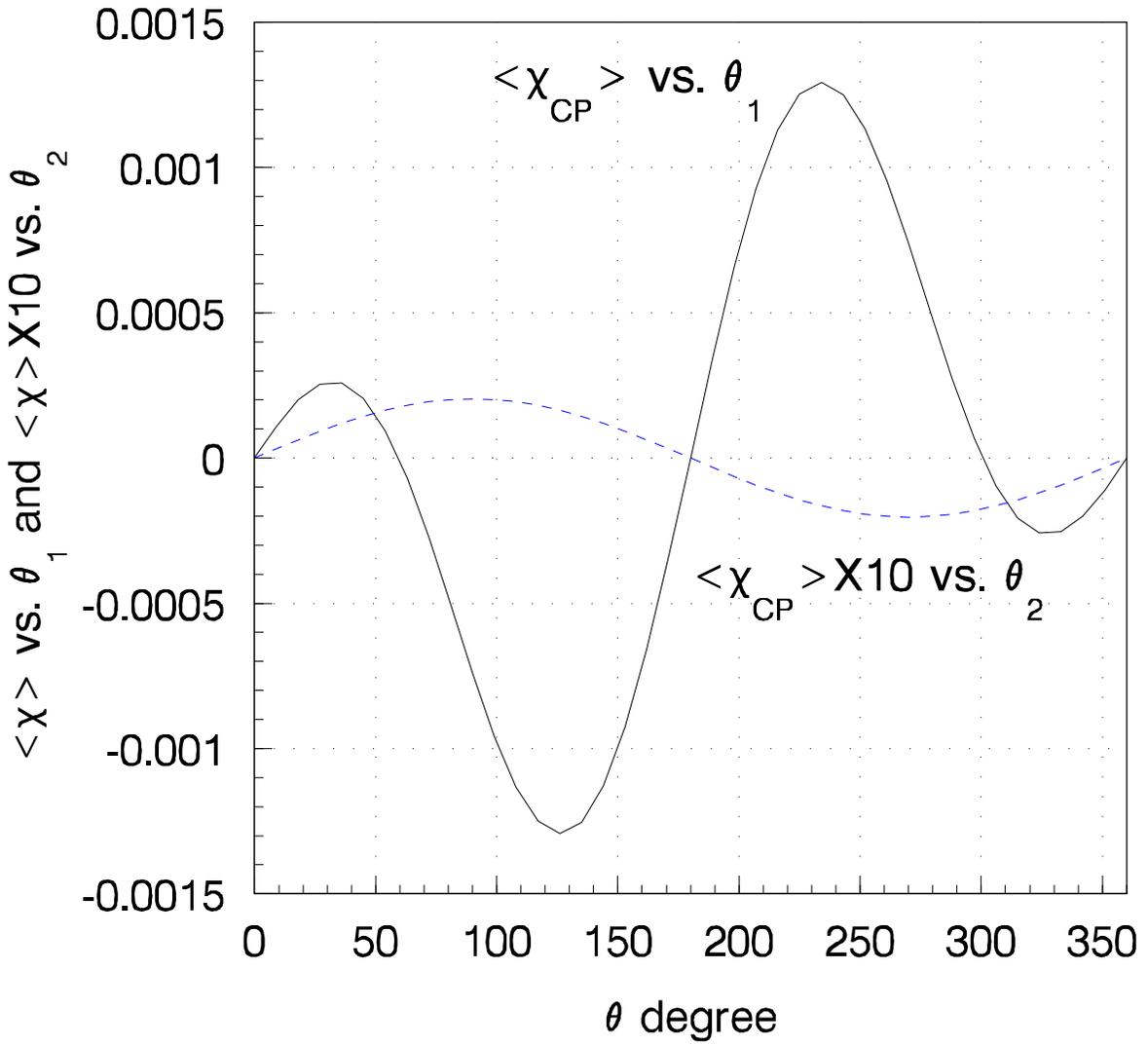
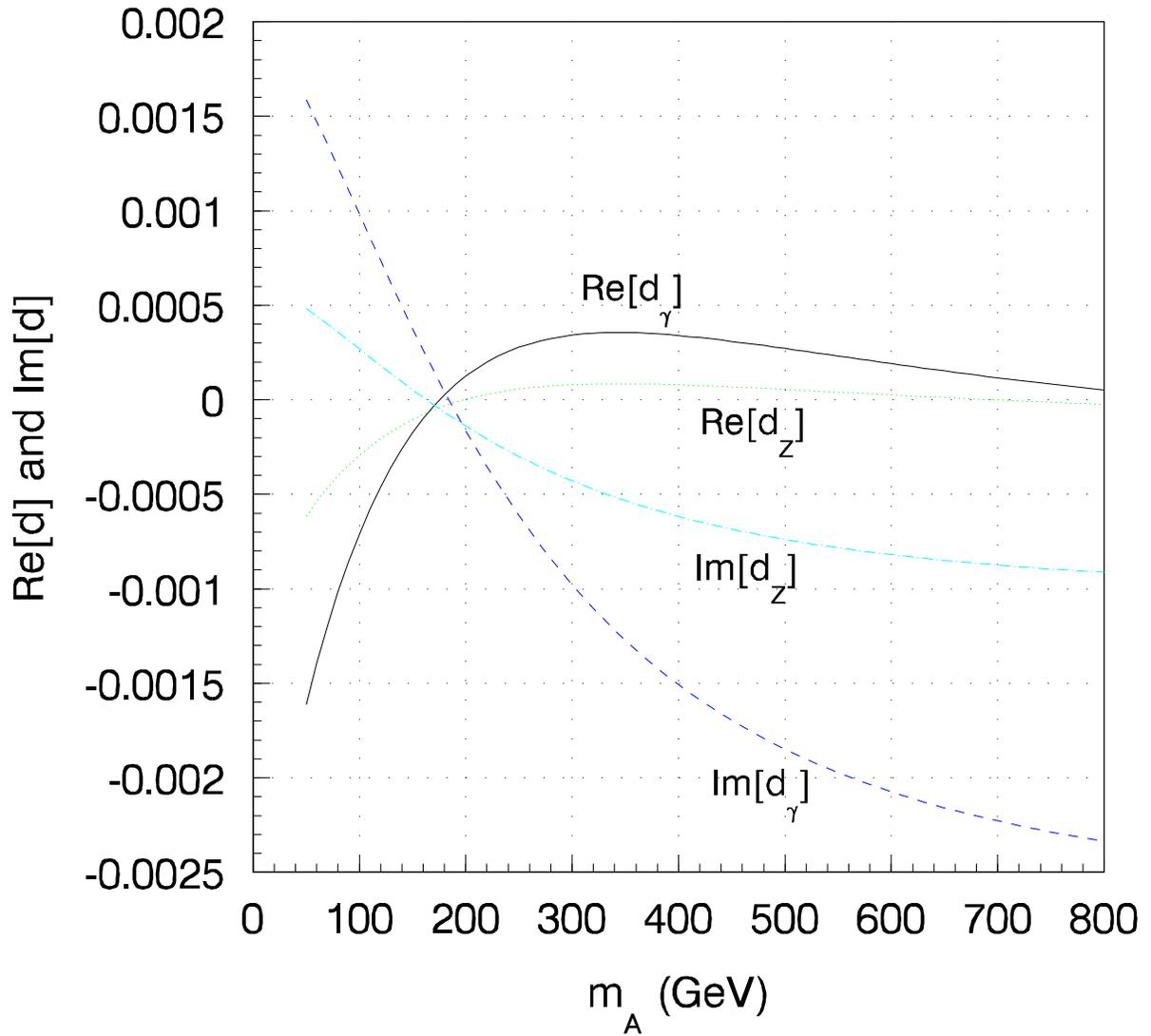
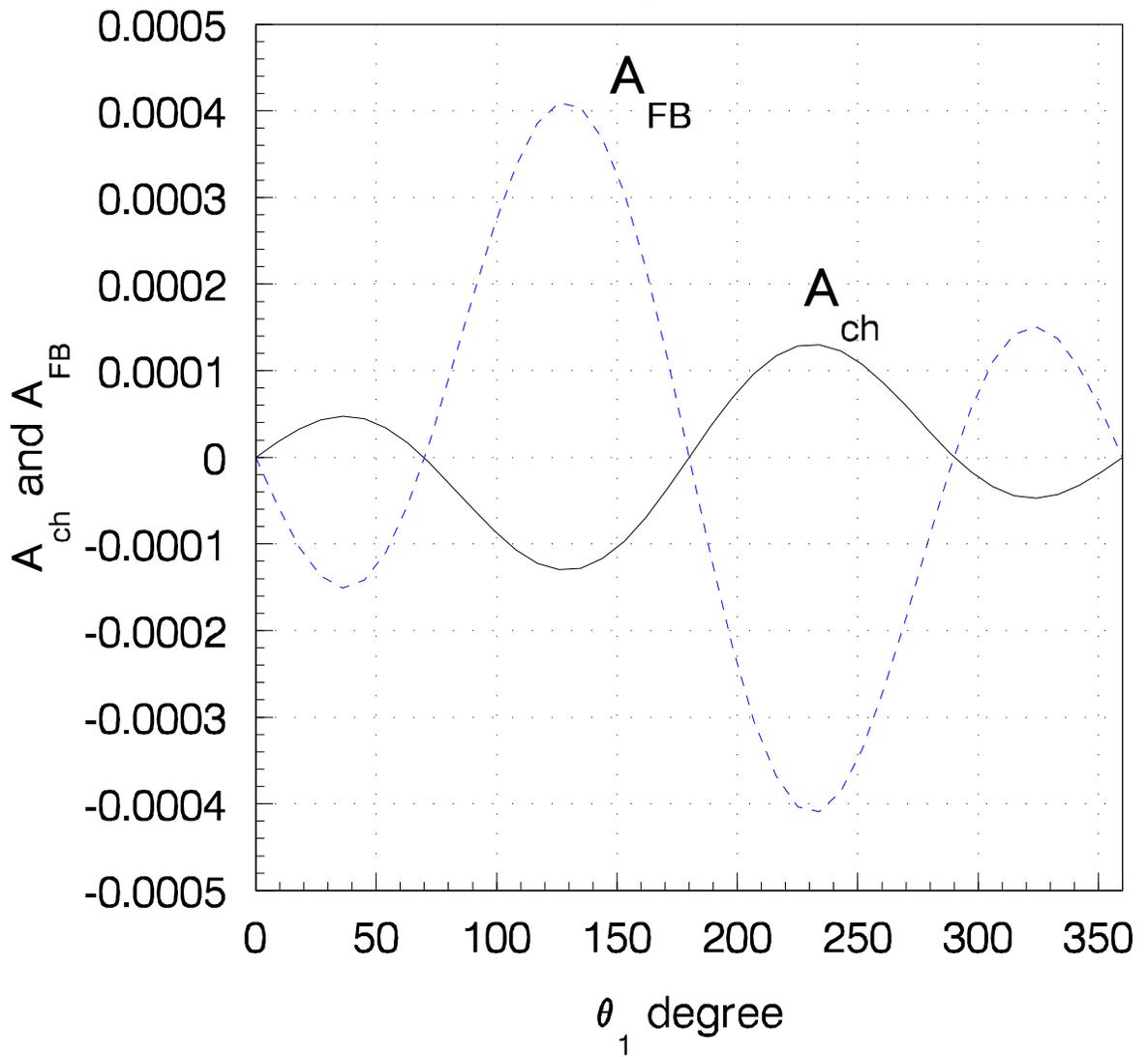
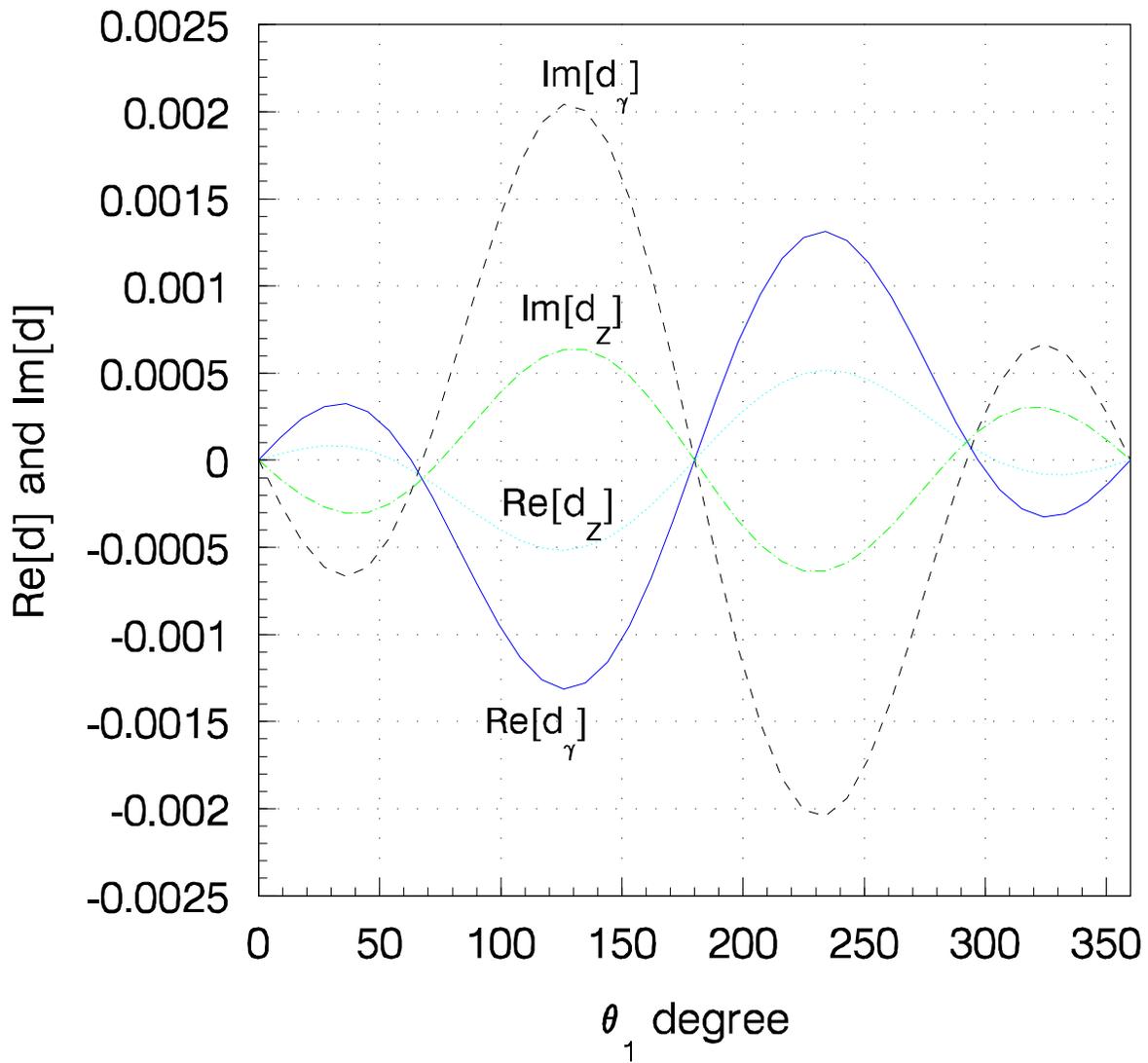
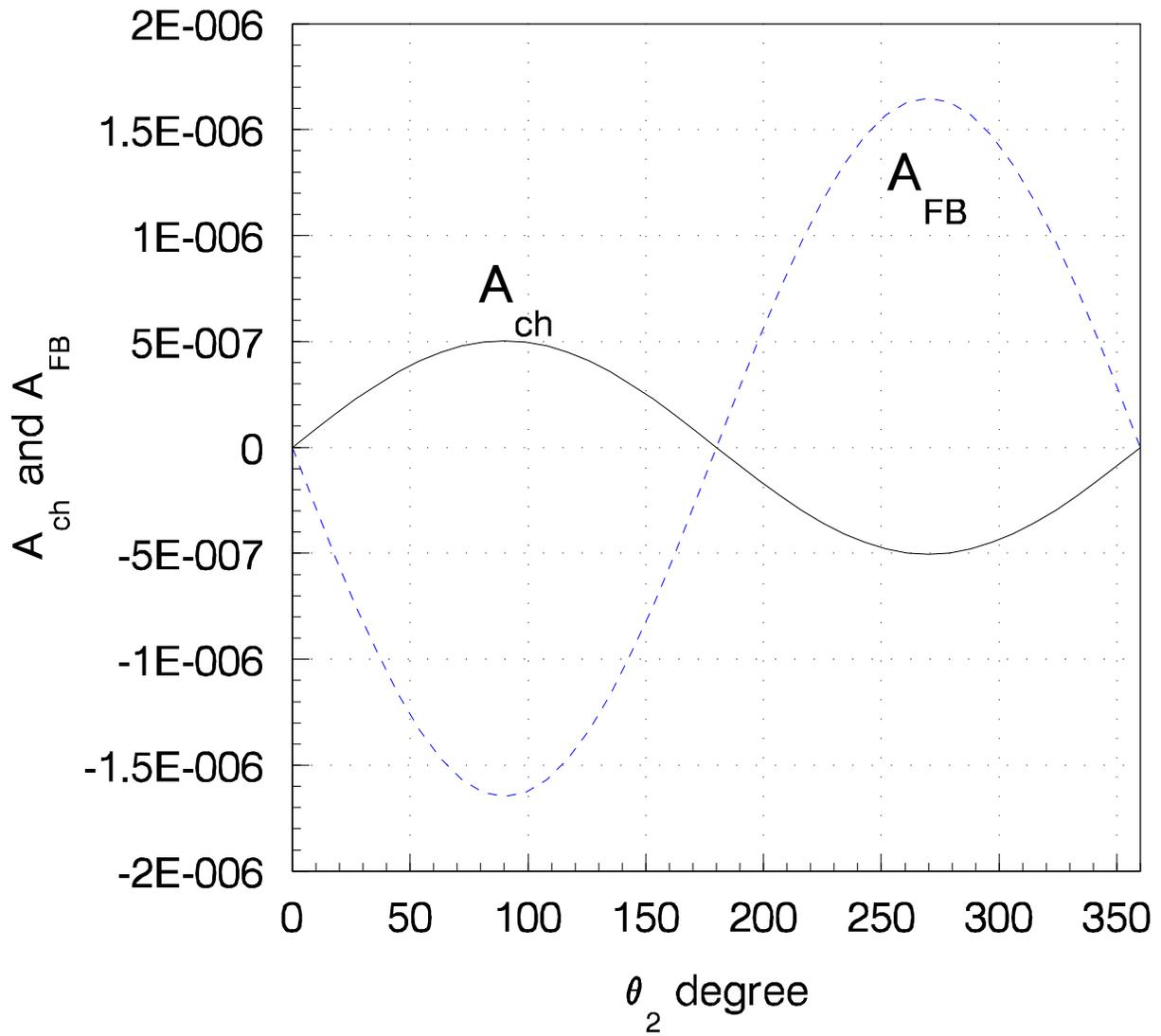
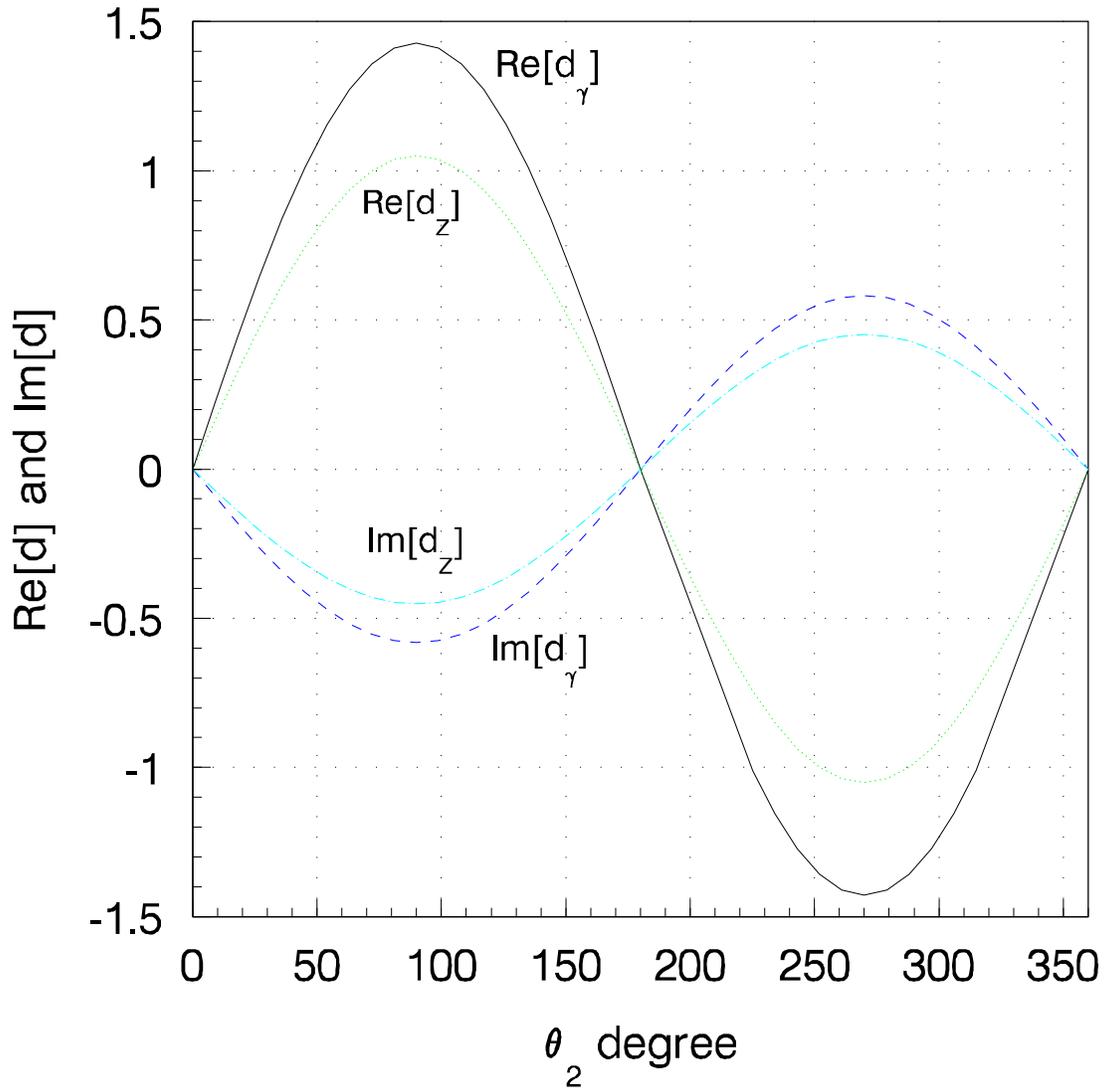
\end{document}